\documentclass[proceedings]{JHEP3}

\PrHEP{PrHEP hep2001}
\conference{International Europhysics Conference on HEP}

\usepackage{epsfig}
\usepackage{amsmath}
\usepackage{amssymb}

\title{A New Model-Independent Analysis of $B \rightarrow X_s \gamma$\\
 in  Supersymmetry
\thanks{Work partially supported by Schweizerischer Nationalfonds.}}

\author{Thomas Besmer\\
Institute for Theoretical Physics, University of Zurich, CH--8057 Zurich, Switzerland}              
\author{Christoph Greub\\
Insitute for Theoretical Physics, University of Berne, CH--3012 Berne, Switzerland}  
\author{\speaker{Tobias Hurth}\\
CERN, Theory Division, CH--1211 Geneva 23, Switzerland}

\abstract{
Using the example of the rare decay $B \rightarrow X_s \gamma$, 
we analyse the importance of interference effects for the bounds on the 
parameters in the squark mass matrices within the unconstrained MSSM.
In former model-independent analyses no 
correlations between the different sources of flavour violation 
were taken into account.
In our new analysis we 
include the contributions from charged Higgs bosons, charginos
and neutralinos and their interference effects and, even more important,
the effects that appear  when several flavour violating parameters,
i.e. several off-diagonal elements in the squark mass matrices, 
are switched on simultaneously. 
We derive new bounds  on certain  off-diagonal elements of 
the squark-mass matrix which are in general one order of 
magnitude weaker than the previous bounds.\\
CERN-TH/2001-331, \,\, BUTP-01/19, \,\, ZH-TH-42/01}

\begin{document}
%
%
  \section{Introduction}
    \label{sec:intro}
Flavour changing neutral current (FCNC) processes
provide crucial guidelines for supersymmetry model building. 
Besides
the Cabibbo--Kobayashi--Maskawa (CKM)-induced contributions,
there are  generic supersymmetric  contributions induced by 
flavour mixing  in the squark mass matrices within 
the so-called unconstrained minimal supersymmetric standard model
(MSSM).
The structure of the MSSM does not explain the suppression of 
FCNC processes, which is observed in experiments;
this is the crucial point of the well-known supersymmetric flavour 
problem.

Among neutral flavour transitions involving the third generation, 
the rare decay $B \rightarrow X_s \gamma$
is at present the most important one~\cite{Review}, as it is
the only inclusive mode that is already 
measured~\cite{BSGMEASURE} and that already provides 
theoretically clean and rather stringent constraints on 
the parameter space of various extensions of the 
SM~\cite{NLLBEYOND}.
Although the theoretical SM prediction,  up to  next-to-leading 
logarithmic (NLL) precision~\cite{NLL}
for its branching ratio,
is in agreement with the experimental data, it is still possible 
that the rare decay $B \rightarrow X_s \gamma$  leads to the first 
evidence of new physics by a significant deviation from the 
SM prediction, for example in the observables concerning direct
CP violation~\cite{CP}.

The decay  $B \rightarrow X_s \gamma$ is sensitive to the 
mechanism of supersymmetry breaking because,  
in the limit of exact supersymmetry, the decay rate would
be just zero, ${\cal B}(B \to X_s \gamma) = 0$.  
Flavour violation thus originates
from the interplay  between the dynamics of flavour and the mechanism of  
supersymmetry breaking and FCNC processes 
may contribute to the question of 
which mechanism ultimately breaks the supersymmetry.

Former analyses in the unconstrained MSSM neglected QCD corrections
and only used  the gluino contribution to saturate the experimental bounds. 
Moreover, no correlations between different sources of flavour 
violation were taken into account. 
In this way, one arrived at  `order-of-magnitude bounds' on the 
soft parameters~\cite{GGMS,DNW,HKT}.
In~\cite{BGHW}, the sensitivity of the bounds on the down squark
mass matrix to radiative QCD corrections was analysed, 
including the SM and the gluino contributions.
In the new analysis~\cite{NEW} we present here, we 
include the contributions from charged Higgs bosons, charginos
and neutralinos and their interference effects and, even more important,
the effects that result when several flavour violating parameters,
i.e. several off-diagonal elements in the squark mass matrices, 
are switched on simultaneously. 

\section{Phenomenological Analysis}
    \label{sec:analysis}
To understand the sources of flavour violation that may be present in
supersymmetric models, in addition to those enclosed in the CKM matrix,
one has to consider the contributions to the squark mass matrices
\begin{equation}
{\cal M}_f^2 \equiv  \left( \begin{array}{cc}
  m^2_{\,f,\,LL} +F_{f\,LL} +D_{f\,LL}           & 
                 \left(m_{\,f,\,LR}^2\right) + F_{f\,LR} 
                                                     \\[1.01ex]
 \left(m_{\,f,\,LR}^{2}\right)^{\dagger} + F_{f\,RL} &
             \ \ m^2_{\,f,\,RR} + F_{f\,RR} +D_{f\,RR}                
 \end{array} \right) \,,
\label{massmatrixd}
\end{equation}
where $f$ stands for up- or down-type squarks.
We recall that the matrices $m_{u,LL}$ and $m_{d,LL}$  cannot be specified independently;
$SU(2)_L$ gauge invariance implies that 
$m_{u,LL} =  K m_{d,LL} K^\dagger$, where $K$ is the CKM matrix. 
In the super-CKM basis, where the quark mass matrices are diagonal 
and the squarks are rotated in parallel to their superpartners,
the $F$ terms  from the superpotential and the $D$ terms 
turn out to be diagonal 
$3 \times 3$ submatrices of the 
$6 \times 6$
mass matrices ${\cal M}^2_f$. This is in general not true 
for the additional terms $m^2_{f}$, originating from  the soft 
supersymmetry breaking potential. 
Because all neutral gaugino couplings are flavour diagonal
in the super CKM basis, the 
gluino contributions to the
decay $B \rightarrow X_s \gamma$ are induced by the off-diagonal
elements of the soft terms 
$m^2_{f,LL}$, $m^2_{f,RR}$, $m^2_{f,RL}$.
Since there are different contributions to this
decay, with different numerical impact on its rate, some of these
flavour violating terms may turn out to be poorly constrained. Thus, 
given the generality of such a calculation, it is 
convenient to rely on the mass eigenstate 
formalism, which remains valid even when some of the intergenerational mixing 
elements are large, and not to use the approximate mass insertion 
method (MIA), where the off-diagonal squark mass matrix elements are taken 
to be small and their higher powers neglected. In the latter approach
the reliability of the approximation can be checked only a posteriori.

As a first step, it is convenient to select {\it one}  possible 
source of flavour violation in the squark sector at a time and
assume that all the remaining ones are vanishing. 
It should be stressed that one already excludes any kind
of interference effects between different sources of flavour
violation in this way. 
Following
ref.~\cite{GGMS}, all diagonal entries in 
$m^2_{\,d,\,LL}$, $m^2_{\,d,\,RR}$, and $m^2_{\,u,\,RR}$
are set equal and their common value is denoted by
$m_{\tilde{q}}^2$.  The branching ratio can then be studied as a
function of 
\begin{equation} 
\delta_{LL,ij} = \frac{(m^2_{\,d,\,LL})_{ij}}{m^2_{\tilde{q}}}\,, 
\hspace{0.1truecm} \qquad
\delta_{RR,ij} = \frac{(m^2_{\,d,\,RR})_{ij}}{m^2_{\tilde{q}}}\,, 
\hspace{0.1truecm} \qquad
\delta_{LR,ij} = \frac{(m^2_{\,d,\,LR})_{ij}}{m^2_{\tilde{q}}}\, 
\,\,\,\,\, (i \ne j).
\label{deltadefa}
\end{equation}
We find  that only those parameters get stringently
bounded by $B \to X_s \gamma$, which can generate
contributions to the two five-dimensional 
gluino-induced dipole operators, 
namely $\delta_{d,LR,23}$ and $\delta_{d,RL,23}$.
The two corresponding 
operators are connected by chirality and are denoted 
by ${\cal O}_{7\tilde{g},\tilde{g}}$ and  $\hat{{\cal O}}_{7\tilde{g},\tilde{g}}$ in the following.
As the gluino yields, intrinsically, the dominant contribution by far, 
we also find that the bounds
on $\delta_{d,LR,23}$ and $\delta_{d,RL,23}$ are only marginally modified
by chargino, neutralino and charged Higgs boson contributions.

\begin{figure}[t]
    \begin{center}
    \leavevmode
    \includegraphics[height=5cm,bb=20 145 560 586]{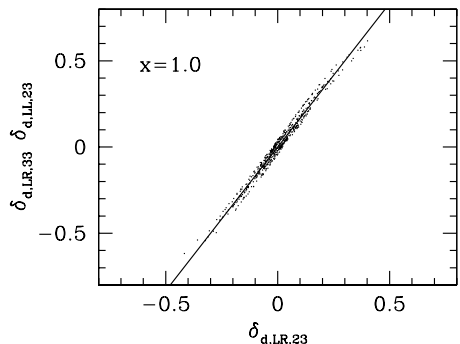}
    \vspace{-4ex}
    \caption[f1]{The parameters $\delta_{d,LR,23}$ and 
              $\delta_{d,LR,33} \cdot \delta_{d,LL,23}$, 
              which are compatible
              with the data on $B \to X_s \gamma$,  are shown by dots.
              Values lying on the solid line lead to a vanishing contribution
              of the five-dimensional magnetic dipole operator ${\cal O}_{7\tilde{g},\tilde{g}}$ 
               in the MIA. See text.} 
\vspace{-4ex}
    \label{fig:linie}
    \end{center}
\end{figure}

\begin{figure}[t]
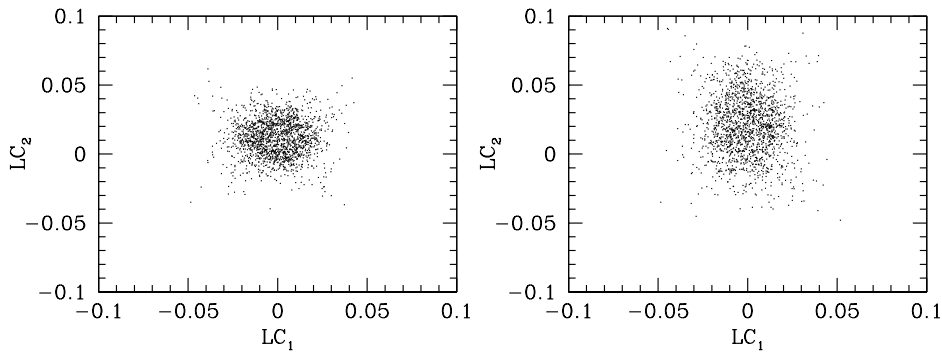

    \begin{center}
    \leavevmode
    \includegraphics[height=5cm,bb= 20 145 560 586]{bild5oben.epsi}
    \includegraphics[height=5cm,bb= 20 145 560 586]{bild5unten.epsi}
    \vspace{-4ex}
    \caption[f1]{Contours in the $LC_1$--$LC_2$ plane with
           $\delta_{d,LR,23}$, $\delta_{d,RL,23}$, 
           $\delta_{d,LL,23}$, $\delta_{d,RR,23}$, 
           $\delta_{d,LR,22}$, $\delta_{d,LR,33}$,
           $\delta_{u,LR,23}$, 
           $\delta_{u,RL,23}$, 
           $\delta_{u,LL,23}$, 
           $\delta_{u,RR,23}$, and
           $\delta_{u,LR,22}$
           all non-vanishing. In the left frame,
           we consider only SM and gluino contributions whereas in the 
           right frame
           we also include chargino, charged Higgs boson and 
           neutralino contributions.}
 \vspace{-6ex}
    \label{fig:lc1lc2}
   \end{center} 
\end{figure}

In the second part of our analysis,  we investigate whether 
the obtained bounds remain stable 
if {\it all} off-diagonal elements, which induce the 
decay $B \rightarrow X_s \gamma$, are varied simultaneously.
For our analysis we explore various scenarios that are characterized
by the values of the parameters $\mu, M_{H^-},   \tan \beta,  M_{\rm{susy}},
  m_{\tilde{g}}$.
We regard this as reasonable, because we expect that
these input parameters, which are unrelated to flavour physics,
will be fixed from flavour conserving observables in the next generations
of high energy experiments (provided low energy SUSY exists).
We note that the common SUSY scale, $M_{\rm{susy}}$,
fixes in our scenarios the general soft squark mass scale $m_{\tilde{q}}$ 
(see (\ref{deltadefa})) 
and the first diagonal element of the chargino mass matrix $M_2$. 
The mass of the charged Higgs boson  $M_{H^-}$ are fixed to be
$M_{H^-} = 300 \,GeV$.
We also allow for a non-degeneracy of the 
diagonal elements in the
matrices $m^2_{\,d,\,LL}$, $m^2_{\,d,\,RR}$, and $m^2_{\,u,\,RR}$.
To implement this, we define 
$\delta$-quantities in addition to those in eq.
(\ref{deltadefa}), which parametrize this non-degeneracy:
$\delta_{f,LL,ii} = ((m^2_{\,f,\,LL})_{ii} - m^2_{\tilde{q}})
/(m^2_{\tilde{q}})$; and analogously for $\delta_{f,RR,ii}$.
In our Monte Carlo analysis
these diagonal $\delta$-parameters are varied
in the interval $[-0.2,0.2]$. On the other hand, the
off-diagonal ones
(in eq. (\ref{deltadefa}))  are varied
in the interval $[-0.5,0.5]$, by use of a Monte Carlo program. There are,
however, two exceptions. First,  we do not vary  
those off-diagonal $\delta$'s with an index $1$; the latter $\delta$'s
we set to zero, since they are severely constrained by kaon decays
(see for example \cite{GGMS}). 
Second, also $(m^2_{u,LR})_{33}$ is not varied, but fixed
such that the mass of the lightest neutral Higgs boson gets heavy enough to be
compatible with experimental bounds \cite{heinemeyer}.
We  plot those events, 
corresponding to  $2.0\times 10^{-4}\leq BR(B \rightarrow X_s \gamma) 
\leq 4.5\times 10^{-4}$, which is the range allowed
by the CLEO measurement. 
\begin{figure}[t]
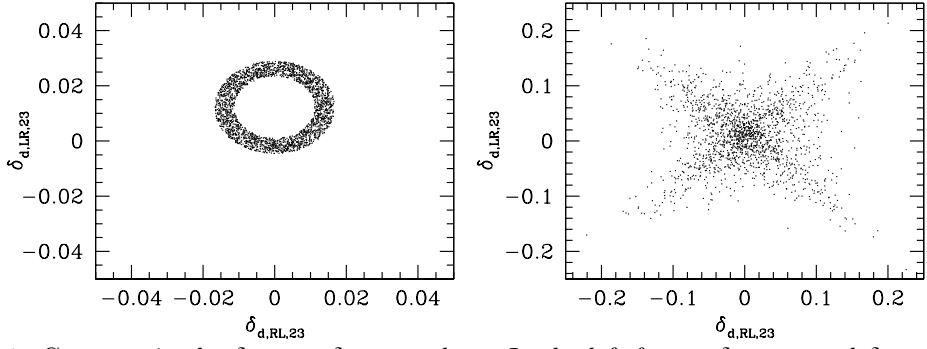

    \begin{center}
    \leavevmode
    \includegraphics[height=5cm,bb= 20 145 560 586]{bild3oben.epsi}
    \includegraphics[height=5cm,bb= 20 145 560 586]{bild4.epsi}
    \vspace{-4ex}
    \caption[f1]{Contours in the $\delta_{d,LR,23}$-$\delta_{d,RL,23}$ plane.
             In the left frame, $\delta_{d,LR,23}$ and 
             $\delta_{d,RL,23}$ are the
             only flavour violating parameters. In the right frame,
           $\delta_{d,LR,23}$, $\delta_{d,RL,23}$, $\delta_{d,LL,23}$ and 
           $\delta_{d,RR,23}$ are flavour violating parameters;
           $\delta_{d,LR,22}$, $\delta_{d,LR,33}$ are also non-zero.
             We only
             consider SM and gluino contributions and the other 
             parameters are $\mu=300\,GeV$, $\tan\beta=10$, 
             $M_{\rm{susy}}=500\,GeV$, $x=1$.}
\vspace{-8ex}
    \label{fig:ad23ad32}
    \end{center}
\end{figure}
We start with the following parameter set:  $\mu=300\,GeV$,
$M_{H^-}=300\, GeV$, $\tan\beta=10$,
$M_{\rm{susy}}=500\,GeV$, 
$x= m_{\tilde{g}}^2 \, / \, M_{\rm{susy}}^2 = 1$ and
$X_t=750\,GeV$. 
In fig. \ref{fig:ad23ad32}, we only consider SM and gluino 
contributions. In the left frame  we  present the constraints  on
$\delta_{d,LR,23}$ and $\delta_{d,RL,23}$ when these are 
the only flavour-violating soft parameters;
the diagonal $\delta$-parameters defined above
are also switched off.
As expected, stringent constraints are obtained.
The hole inside the dotted area represents values of
$\delta_{d,LR,23}$ and $\delta_{d,RL,23}$
for which the branching ratio is too small to be compatible with
the measurements.
In the right frame we now investigate interference 
effects from different sources
of flavour violation, where we allow for non-zero 
 $\delta_{d,LR,23}$, $\delta_{d,RL,23}$, $\delta_{d,LL,23}$, 
$\delta_{d,RR,23}$, $\delta_{d,LR,22}$, and $\delta_{d,LR,33}$. 
All these parameters are varied between $\pm 0.5$. As can be seen, 
the bounds on $\delta_{d,LR,23}$ and 
$\delta_{d,RL,23}$ get destroyed dramatically (note the different scale in
the two plots). The reason for this is that
there are now new contributions to the five-dimensional dipole
operators. As an example,   
the combined effect of $\delta_{d,LR,33}$ and
$\delta_{d,LL,23}$ leads  to a contribution to the Wilson coefficient
of the gluino-induced  magnetic dipole operator 
${\cal O}_{7\tilde{g},\tilde{g}}$. The sign of
this contribution can be different from the one generated by
$\delta_{d,LR,23}$. As a consequence, the bound on $\delta_{d,LR,23}$
gets weakened. To illustrate this more quantitatively, we assume
for the moment that   there are only these
two sources that  can generate  ${\cal O}_{7\tilde{g},\tilde{g}}$,
i.e.  we switch off the other $\delta$-quantities.
If $\delta_{d,LR,23}$ is larger than the individual bound from the
first part of the analysis, it is necessary that the product
of  $\delta_{d,LR,33}$ and $\delta_{d,LL,23}$ also be relatively
large; only in this case can the two sources lead to a branching
ratio compatible with experiment. This feature is illustrated 
in fig. \ref{fig:linie}; only values of  $\delta_{d,LR,23}$
and values of  $\delta_{d,LR,33} \cdot \delta_{d,LL,23}$ that are 
strongly correlated lead to an acceptable branching ratio. 
As clearly visible from fig. \ref{fig:linie}, the correlation
between the two sources for  ${\cal O}_{7\tilde{g},\tilde{g}}$ 
is essentially linear. This implies that the linear combination
$com :=  \delta_{d,LR,23} + f \, \, \delta_{d,LR,33} \cdot \delta_{d,LL,23}$
gets constrained if $f$ is chosen appropriately. Stated differently, 
the Wilson coefficient of the five-dimensional magnetic dipole operator 
is essentially proportional to the combination $com$ defined above.
This implies in turn,  that for the values of the parameters we are using 
at the moment,  
the Wilson coefficient is well approximated by its double 
mass insertion expression. Thus, the coefficient $f$  can be
fixed analytically (see~\cite{NEW});
the numerical values of $f(x)$ for some values of $x$ read $0.74$ for $x=0.3$,
$0.68$ for $x=0.5$,  $0.60$ for $x=1.0$ and $0.52$ for $x=2.0$, respectively.
The solid line in fig.  \ref{fig:linie} represents 
pairs ($\delta_{d,LR,23}$, $\delta_{d,LR,33} \cdot \delta_{d,LL,23}$)
for which the combination $com$ is zero. The points
scattered around this line therefore represent Monte Carlo events for which 
this combination is small.
We now turn back to the scenario of fig. \ref{fig:ad23ad32},
in which all the parameters  
$\delta_{d,LR,23}$, $\delta_{d,RL,23}$, $\delta_{d,LL,23}$,
$\delta_{d,RR,23}$, $\delta_{d,LR,22}$, $\delta_{d,LR,33}$
are varied simultaneously. In this case, the linear combinations
\begin{eqnarray}
LC_1 & = & \delta_{d,RL,23}+f(x)\delta_{d,RR,23}\cdot\delta_{d,RL,33} 
           +f(x)\delta_{d,RL,22}\cdot\delta_{d,LL,23},\nonumber\\ 
LC_2 & = & \delta_{d,LR,23}+f(x)\delta_{d,LR,22}\cdot\delta_{d,RR,23}
           +f(x)\delta_{d,LL,23}\cdot\delta_{d,LR,33}, 
\label{deflc1lc2}
\end{eqnarray}
are expected to get constrained. 
In fig.  \ref{fig:lc1lc2} we show the allowed region for $LC_1$ and $LC_2$. 
There, we allow all 
non-diagonal $\delta$-parameters to vary between $\pm 0.5$. 
In addition, we also allow for non-equal diagonal soft entries, by varying
the parameters $\delta_{f,LL,ii}$ and $\delta_{f,RR,ii}$ between   
$\pm 0.2$. With the latter choice
we still guarantee the hierarchy between diagonal and off-diagonal entries,
but we get rid of the unnatural assumption of degenerate diagonal entries.
In the left frame, we 
include only the SM and  gluino contributions. We find that 
the linear combinations $LC_1$ and $LC_2$ indeed get stringently bounded.
In the right frame of fig. \ref{fig:lc1lc2} we test the resistance of 
these bounds 
when the additional contributions (i.e., 
those from charginos, charged Higgs bosons and neutralinos) are turned 
on. In this case also 
$\delta_{u,LR,23}$, $\delta_{u,RL,23}$,  $\delta_{u,LL,23}$,
$\delta_{u,RR,23}$ and 
$\delta_{u,LR,22}$ are varied in the range $\pm 0.5$.
We find that the bound on $LC_1$ remains unchanged, while the one
on $LC_2$ gets somewhat weakened. This feature is expected, because
charginos and charged Higgs bosons contribute to
the unprimed operator. 

At this point we should stress
that these plots were obtained by choosing the renormalization
scale $\mu_b=4.8\,GeV$ and by requiring all squark masses to be
larger than $150 \,GeV$. We  checked that 
the bounds on $LC_1$ and
$LC_2$ remain practically unchanged when the renormalization
scale is varied between $2.4\, GeV$ and $9.6\, GeV$; they are also insensitive
to the value of the required minimal squark mass, as we found by 
changing 
$m_{\rm{squark\,min}}$ from $150\, GeV$ to $100\, GeV$ or $250\, GeV$. 
Moreover, we also checked 
whether the restriction to  the $\mu=+300$ $GeV$ scenario is 
too severe: we redid the complete analysis 
for $\mu=-300\,GeV$ and confirmed that there are  
no differences between the results with 
these two choices.

There is a remark in order. If we got rid of the
hierarchy of diagonal and off-diagonal entries in the squark mass matrices, 
stringent bounds on the simple combinations $LC_1$ and $LC_2$ 
certainly would no longer exist, simply because there would then be more 
contributions to the five-dimensional operators of similar 
magnitude.
In this case, however, the $full$ Wilson coefficients
of the five-dimensional operators  
 would still be stringently constrained by the experimental
data on $B \to X_s \gamma$. Unfortunately, in this case not much information
can be extracted
for the individual soft parameters or simple combinations thereof.

Finally, we extend our analysis to other values of the input parameters.
We analyse the bounds on the soft parameters within the following 
parameter sets: $M_{\rm{susy}}$ $=$ $300\,GeV$,
$500\, GeV$, $1000\, GeV$.
For $\tan\beta$ we explore the values: $\tan \beta = 2,\, 10,\,  30,\,  50.$
Furthermore, the gluino mass $m_{\tilde{g}}$ is varied over the values
$ x = m_{\tilde{g}}^2 \, / \, M_{\rm{susy}}^2 = 0.3\,, \, 0.5\,,\,  1\,,\, 2$. 
Surprisingly, we find that the constraints on $LC_1$ and 
$LC_2$ are completely stable 
over large parts of the parameter space. 
However, the bounds get 
weakened when $\tan \beta$ values as large as 
$50$ are chosen. 
This effect gets  enhanced when the general mass scale
$m_{\tilde{q}}$ in the 
squark mass matrices decreases with the parameter $M_{\rm susy}$.
There are two main reasons why the bounds get weakened in these scenarios.
First, in the large $\tan \beta$ regime, the term 
$(F_{d,LR})_{33}$ gets strongly enhanced because it is proportional  
to $\tan \beta$. 
Particularly, for $\tan \beta = 50$ 
and $M_{{\rm susy}} = 300\, GeV$, the term  $(F_{d,LR})_{33}$ is of the
same magnitude as the diagonal entries of the squark mass matrix. 
Thus,  the contributions to the Wilson coefficients of the five-dimensional
gluino operators (induced by $(F_{d,LR})_{33}$
in combination with
$\delta_{d,LL,23}$ or $\delta_{d,RR,23}$) become important 
enough to weaken the bounds on $LC_1$ and $LC_2$ significantly.
The relative importance of this $F$ term is of course increased  
if the general soft squark mass scale $M_{{\rm susy}}$ is decreased.
Second, within the large $\tan \beta$ regime  
the contributions from charginos get enhanced
and therefore also weaken the bounds on $LC_2$. 
Because the parameter $(F_{d,LR})_{33}$ is actually proportional to the 
product of $\tan \beta$ and $\mu$, we conclude 
from the above findings that the bound on $LC_1$ is  unchanged if we 
increase the value of $\mu$ and decrease the value of $\tan \beta$ so
that the product of both parameters is constant; the bound on
$LC_2$ is then  even stronger,  because the chargino contribution 
is smaller for increasing $\mu$. 

\section{Conclusions}
    \label{sec:conclusions}
Our new  model-independent analysis of the rare decay
$B\rightarrow X_s \gamma$, based on a systematic leading
logarithmic QCD analysis,  mainly explored
the interplay  between the various 
sources of flavour violation and the interference effects of
SM, gluino,
chargino, neutralino, and charged Higgs boson contributions.
In former analyses, no 
correlations between the different sources of flavour violation 
were taken into account.
Unlike previous work, which used the mass insertion approximation, 
we used in our analysis the mass eigenstate formalism, which remains 
valid even when some of the  intergenerational mixing elements are 
large. 
We singled out  two simple combinations 
 of elements of the soft parts of the down squark mass matrices,
which stay  stringently bounded over large parts of the 
supersymmetric parameter space, 
excluding the large $\tan \beta$ and the large $\mu$ regime.
These new bounds are in general one order of magnitude weaker 
than the bound on the single off-diagonal element $\delta_{d,LR,23}$, which 
was derived in previous work \cite{GGMS,Masiero2001}
by neglecting any kind of interference effects. 
It seems that the flavour problem is less severe
in the $B$ system than often stated. 
Thus, it would be interesting to explore also the consequences 
of natural interference effects between different sources
of flavour changes within the kaon sector.

\end{document}